\documentstyle[12pt]{article}
\begin{document}

\begin{center}
{\Large 
Heavy top limit and 
double-logarithmic contributions to Higgs production 
at $m_H^2 / s \ll 1$}

\vskip 50pt
F. Hautmann\\
\vskip 10pt
Physics Department, University of Oregon\\ 
Eugene OR 97403, USA\\ 
\vskip 10pt
and\\  
\vskip 10pt
Institut f{\" u}r Theoretische Physik, Universit{\" a}t Regensburg\\
 D-93040 Regensburg, Germany\\
\vskip 50pt
\noindent{\bf Abstract}
\vskip 20pt
\end{center}

\noindent{ 
 Next-to-next-to-leading order (NNLO) QCD corrections to 
Higgs boson hadroproduction 
have recently been calculated
in the heavy 
top-quark  limit $m_t \to \infty$.  
 The $m_t \to \infty$ limit introduces 
double-logarithmic  corrections in $\ln x$, with  $x 
\equiv m_H^2 / s$.  We  identify 
  these corrections order by order in $\alpha_s$. As an 
  application,   we  derive an analytic 
expression  for the dominant $x \ll 1$  
part of the NNLO coefficient. }

\newpage

Recently, Harlander and Kilgore~\cite{hk02} have 
completed the 
calculation of  the 
next-to-next-to-leading order (NNLO) QCD correction to 
 Higgs boson production in hadron-hadron collisions.  The calculation 
 is done in the heavy top quark limit, in which $m_t \to \infty$ and all 
 other quark masses vanish.  In this limit, the 
  effective Lagrangian 
 of \cite{jellis} is used for  coupling  the Higgs boson 
 to gluons. 
The hard scattering coefficient expanded through NNLO 
has the structure 
\begin{eqnarray} 
\label{coeffexp}
C ( \alpha_s, x, m_H^2 /  \mu^2 )  &=&
c^{(0)} ( x ) + { {\alpha_s } \over \pi } \ 
\left[ c^{(1)} ( x )  + 
{\overline c}^{(1)} ( x ) \ L \right] 
\nonumber\\ 
&+&  
\left( { { \alpha_s } \over \pi } \right)^2 \ 
\left[ c^{(2)} ( x )  + 
{\overline c}^{(2)} ( x ) \ 
L 
+ {\overline {\overline c}}^{(2)} ( x ) \ L^2 
\right] 
+ \, \cdots 
\end{eqnarray} 
where $L \equiv \ln (m_H^2 / \mu^2)$, with 
$m_H$  the Higgs boson mass and $\mu$ the factorization scale;  
 $x = m_H^2 / s$, with 
$s$ the center-of-mass energy. The term in  $c^{(0)}$ is 
the  leading-order term~\cite{loterm}; the term in  
  $c^{(1)}, {\overline c}^{(1)}$  is 
 the next-to-leading-order term~\cite{nloterm}.

Ref.~\cite{hk02} 
presents the result for $c^{(2)}$ 
as an  expansion in powers of $(1 - x)$, 
  $x = 1$ being the production threshold. 
Explicit numerical results are given 
up to a very high order 
in this expansion, $(1-x)^{16}$. 
Once the hard scattering coefficient 
is convoluted with the parton densities, this expansion leads to 
a corresponding  
   expansion for the cross section. 
   This is demonstrated numerically to converge well~\cite{hk02}. 
As discussed    by 
Catani, de~Florian and Grazzini~\cite{cat02} and   
by Harlander and Kilgore~\cite{hk02},  the  
reason for the convergence 
of the $(1 - x)$ expansion  is mainly kinematical, and 
depends on  the 
steeply rising behavior of  parton luminosities for decreasing 
momentum fractions.  
This behavior also explains why    results based only on 
soft~\cite{soft01,soft98} and virtual~\cite{harl00}  
contributions give a reasonably good  
approximation to the full answer.

In this note we 
focus on   the influence of the 
$m_t \to \infty$ approximation   on  the 
 $x \ll  1$  part of the coefficient. 
 It was observed in 
 \cite{ktcch} that a  local 
  Higgs-gluon  coupling 
 gives rise to   double-logarithmic corrections 
  in  $\ln x$.  In this note  
   we  identify these corrections  explicitly order 
  by order in $\alpha_s$.  In particular 
 we obtain an    analytic  result  for  
  the dominant $x  \ll  1$ part  of the  NNLO coefficient.

Although the $x  \ll  1$ behavior   is 
unlikely to  
affect  the convergence of the inclusive cross section at NNLO, it 
may  however  affect  distributions 
associated to less inclusive observables. 
This analysis may thus be useful for further 
evaluations of the hard scattering.

Let us consider the 
  $x  \ll  1$ corrections 
in the framework of 
 the 
effective Lagrangian~\cite{jellis} for the 
Higgs boson coupling to gluons:    
\begin{equation}  
\label{leff}
{\cal L}_{\rm{eff}} =    (G_F \sqrt{2} )^{1/2}  \ K (\alpha_s) \ 
G_{\mu \nu}^a G^{a \mu \nu} \ H
\hspace*{0.3 cm} ,    
\end{equation}
where $K(\alpha_s)=\alpha_s/(12 \pi)+
{\cal O}(\alpha_s^2)$ is the  coefficient function  containing the  
 dependence on the top  quark mass, and is known to order 
$\alpha_s^4$~\cite{chetetal,soft98}. 
  The  $x  \ll  1$ corrections 
  come, in higher order graphs,    from  
integrating two-gluon irreducible amplitudes 
 over the transverse momenta 
$ {\bf k}_1$ and   ${\bf k}_2$ of the gluons that couple to 
the Higgs boson. In  the full theory,  
contributions from $ |{\bf k}_1| ,  |{\bf k}_2| 
\gg m_t $ are suppressed  
by the top quark loop. In  the effective theory,   
large values of   $ |{\bf k}_1| $ and $  |{\bf k}_2|$    are allowed 
all the way up to the  kinematic limit 
  $\sqrt{s}$.   This results~\cite{ktcch} in the 
  perturbative coefficients having 
stronger singularities in the complex $N$ plane  as $N \to 0$ (with $N$ 
the moment variable 
of Eq.~(\ref{ktfact})).
More precisely,    double poles $\alpha_s / N^2$  appear 
order by order 
in $\alpha_s$. 

To determine these contributions, 
recall  the  structure
  of the 
  hard scattering 
  coefficient in the ${\overline{\mbox{MS}}}$ subtraction scheme 
  for $N \to 0$~\cite{ktcch}:  
\begin{eqnarray} 
\label{ktfact}
C_N (\alpha_s,  {m_H^2 /  \mu^2} )  &\equiv& \int_0^1 
dx x^{N-1} 
C ( \alpha_s, x, {m_H^2 /  \mu^2} ) 
\nonumber\\
&=& R_N^2 (\gamma_N) \ (m_H^2 / \mu^2 )^{2 \gamma_N} \ 
h_N (\gamma_N, \gamma_N) \hspace*{0.3 cm} .   
\end{eqnarray}
Here $R_N$ is a  normalization 
factor associated with the 
choice of the ${\overline{\mbox{MS}}}$ scheme~\cite{msr};  
$\gamma_N$ is a known, universal  anomalous dimension,  
with the perturbation expansion    
\begin{equation}  
\label{andim}
\gamma_N = {C_A  \over N }  { \alpha_s \over \pi} + {\cal O} 
\left( {\alpha_s \over N} \right)^4 \hspace*{0.3 cm} ; 
\end{equation}
the function $h_N$ is constructed from the 
matrix element $\hat \sigma$ 
for the off-shell amplitude $g (k_1) + g (k_2) \to H $ 
by taking  the following  integral transform 
\begin{eqnarray}  
\label{hdef}
h_N (\gamma_1 , \gamma_2) &=& 
\gamma_1 \ \gamma_2 \ 
\int {{d^2 {\bf k}_1 } \over { \pi {\bf k}_1^2 } } 
\left( { {\bf k}_1^2 \over m_H^2 } \right)^{\gamma_1} 
\int {{d^2 {\bf k}_2 } \over { \pi {\bf k}_2^2 } } 
\left( { {\bf k}_2^2 \over m_H^2 } \right)^{\gamma_2} 
\nonumber\\ 
&\times& 
\int_0^1 { {dx } \over x} \ x^N \ {\hat \sigma} \left( x, 
{ {\bf k}_1 \over m_H } , { {\bf k}_2 \over m_H } \right) 
\hspace*{0.3 cm} .   
\end{eqnarray}
The 
matrix element ${\hat \sigma}$, although off-shell,  
is defined gauge-invariantly by coupling the gluons with 
eikonal polarizations, as in \cite{ktcch}. It 
is readily calculated using  the effective  
Lagrangian (\ref{leff}): 
\begin{eqnarray}  
\label{sigeff}
{\hat \sigma} \left( x, 
{ {\bf k}_1 \over m_H } , { {\bf k}_2 \over m_H } \right) 
&=&  { { \alpha_s^2 m_H^2 G_F \sqrt{2} } \over { 288 \pi}} \ 
{m_H^4 \over { x^2 {\bf k}_1^2 {\bf k}_2^2 } } \ 
{ { ( {\bf k}_1 \cdot {\bf k}_2 )^2 } \over 
{ \left( m_H^2 + {\bf k}_1^2 + {\bf k}_2^2 \right)^2 } } 
\nonumber\\ 
&\times& 
{1 \over x} \ \delta \left( {1 \over x} - 1 - 
{ {({\bf k}_1 + {\bf k}_2)^2} \over m_H^2 }  \right)  
\hspace*{0.3 cm} .   
\end{eqnarray}

Now consider the function $h_N$.   The 
$x$ integral in Eq.~(\ref{hdef}) can be done 
using the $\delta$-function of Eq.~(\ref{sigeff}): 
\begin{eqnarray}  
\label{sigmom}
\int_0^1 { {dx } \over x} \ x^N \ {\hat \sigma} &=&  
{ { \alpha_s^2 m_H^2 G_F \sqrt{2} } \over { 288 \pi}} 
{m_H^4 \over {  {\bf k}_1^2 {\bf k}_2^2 } } 
\ 
{ { ( {\bf k}_1 \cdot {\bf k}_2 )^2 } \over 
{ \left( m_H^2 + {\bf k}_1^2 + {\bf k}_2^2 \right)^2 } } 
 \nonumber\\ 
&\times& 
{{(m_H^2)^{N-2}} \over 
{ \left[ m_H^2 + ({\bf k}_1 + {\bf k}_2)^2 \right]^{N-2} } } 
\hspace*{0.3 cm} .   
\end{eqnarray}
Notice that 
when $N \to 0$ the matrix element  is unsuppressed  for  
large $ |{\bf k}_1| ,  |{\bf k}_2| $.  This 
is a consequence of 
approximating the Higgs coupling  by a  local vertex, and 
is in contrast with the  
general  case of a finite top-quark mass~\cite{spira95}. 
The behavior at large transverse momenta 
in Eq.~(\ref{sigmom}) 
causes the function $h$  to have  $N=0$ singularities 
not only through its dependence on the anomalous  
dimension (\ref{andim}) but also through  its explicit 
$N$ dependence. 
 Introducing the notation 
\begin{equation} 
\label{xinot}
\xi_1 =  {\bf k}_1^2 / m_H^2 
\hspace*{0.3 cm} , \hspace*{0.7 cm} 
\xi_2 =  {\bf k}_2^2 / m_H^2 \hspace*{0.3 cm}  
\end{equation}
and performing the angular integral in 
 Eq.~(\ref{hdef}) we get  
\begin{eqnarray}  
\label{hhyper}
&& h_N (\gamma_1 , \gamma_2)  = 
{ { \alpha_s^2 m_H^2 G_F \sqrt{2} } \over { 288 \pi}}  
\ \gamma_1 \ \gamma_2 \ \int_0^{\infty} d \xi_1 \ \xi_1^{\gamma_1 - 1} 
\ \int_0^{\infty} d \xi_2 \ \xi_2^{\gamma_2 - 1} 
\nonumber\\ 
&\times& \left\{  {1 \over {2 \ (1 + \xi_1 + \xi_2)^N }} 
\;  { _2F_1}  \left( { {N-1} \over 2} ,{ {N-2} \over 2}  , 2 , 
{ { 4 \xi_1 \xi_2} \over {  (1 + \xi_1 + \xi_2)^2}} \right) 
\right. 
\nonumber\\ 
&+& \left.  
{{\xi_1 \xi_2 \ (N-1) (N-2)} \over { 2 \  (1 + \xi_1 + \xi_2)^{2+N} }} 
\; {_2F_1} \left( { {N} \over 2} ,{ {N+1} \over 2}  , 3 , 
{ { 4 \xi_1 \xi_2} \over { (1 + \xi_1 + \xi_2)^2}} \right)
\right\}  \hspace*{0.2 cm} ,
\end{eqnarray}
where ${_2F_1}$ is the hypergeometric function.  
The integration 
 region of large  $\xi_1$  and $\xi_2$ in Eq.~(\ref{hhyper}) 
gives rise to a pole at $\gamma_1 + \gamma_2 = N$. 
This pole is the origin of the double-logarithmic  terms. 

To 
approximate  $h$ near $N=0$ 
taking the contribution of this pole into account, we 
expand the hypergeometric functions in Eq.~(\ref{hhyper}) 
to the lowest order around $N=0$, 
\begin{equation} 
\label{hypexpan1}
_2F_1 \left( { {N-1} \over 2} ,{ {N-2} \over 2}  , 2 , 
{ { 4 \xi_1 \xi_2} \over { (1 + \xi_1 + \xi_2)^2}} \right) 
\simeq 1 + { {  \xi_1 \xi_2} \over { (1 + \xi_1 + \xi_2)^2}} + 
{\cal O } (N) \hspace*{0.3 cm} ,
\end{equation}
\begin{equation} 
\label{hypexpan2}
(N-1) (N-2) \  _2F_1 \left( { {N} \over 2} ,{ {N+1} \over 2}  , 3 , 
{ { 4 \xi_1 \xi_2} \over { (1 + \xi_1 + \xi_2)^2}} \right) 
\simeq 2 + 
{\cal O } (N) \hspace*{0.3 cm} ,
\end{equation}
but retain the $N$ dependence in the factor $(1 + \xi_1 + \xi_2)^{-N}$. 
The integrations in $\xi_1$ and  $\xi_2$  
can now be carried out simply 
in terms of Euler $\Gamma$ functions:  
\begin{eqnarray} 
\label{hn0}
 h_N (\gamma_1 , \gamma_2) &\simeq& 
{ { \alpha_s^2 m_H^2 G_F} \over { 288 \pi \sqrt{2} }} 
\  \Gamma(1+\gamma_1) \Gamma(1+\gamma_2) 
\Gamma(N-\gamma_1-\gamma_2) 
\nonumber\\ 
&\times& \left[ {1 \over {\Gamma(N)}} + 
3 \gamma_1 \gamma_2 \ {1 \over {\Gamma(N+2)}} \right] \hspace*{0.3 cm} , 
\hspace*{0.6 cm} N \ll 1   \hspace*{0.2 cm} . 
\end{eqnarray}

We can now expand $h$ for small 
anomalous dimensions, setting $\gamma_1 = \gamma_2 
= \gamma_N$, 
with $\gamma_N$ given in Eq.~(\ref{andim}), 
and determine 
the leading $N \ll 1$ part of the 
 perturbative coefficients through the second order: 
\begin{eqnarray}  
\label{hpert}
h_N (\gamma_N , \gamma_N) &\simeq& 
{ { \alpha_s^2 m_H^2 G_F  } \over { 288 \pi \sqrt{2}}} 
\\
&\times& \ 
\left( 1 + 2  \gamma_N {1 \over N} + 4 \gamma_N^2 {1 \over N^2} + 
\ \cdots 
\right) \hspace*{0.7 cm} ( \gamma_N \ll 1 \hspace*{0.1 cm} , 
\hspace*{0.1 cm} N \ll 1 ) 
\nonumber\\
\nonumber
\end{eqnarray}
Via Eq.~(\ref{andim}), Eq.~(\ref{hpert}) gives 
 the double-pole terms 
$\alpha_s / N^2$  order by order in perturbation theory. 
These terms are  generated by  the pole $(N-2 \gamma_N)^{-1}$ 
in $h$, see Eq.~(\ref{hn0}).

We are now in a position to find the NNLO correction 
to the gluon-gluon coefficient for 
$x \ll 1$. 
Higher order contributions to $K(\alpha_s)$~\cite{chetetal} 
in Eq.~(\ref{leff}) 
are subleading 
for $x \ll 1$. 
The factor $R_N$ in Eq.~(\ref{ktfact}) 
is a single-logarithmic factor~\cite{msr} that starts to 
contribute one order higher than 
NNLO: 
\begin{equation} 
\label{rn3}
 R_N = 1 + {\cal O} (\alpha_s / N)^3 \hspace*{0.3 cm} . 
\end{equation}
 The $(m_H^2 / \mu^2)^{2 \gamma_N}$ factor in Eq.~(\ref{ktfact}) 
is to be expanded for small $\gamma_N$, as has been done with $h$. 
Then, by taking an inverse transform  from moment space 
  to $x$-space, we get  
\begin{eqnarray} 
\label{cgg}
 C_{g g} ( \alpha_s, x, {m_H^2 /  \mu^2} ) 
&=& 
{ { \alpha_s^2 m_H^2 G_F} \over { 288 \pi \sqrt{2} }}  
\left[ \delta (1 - x) + 
{ {\alpha_s }\over \pi } \ 
C_A \left( - 2 \ln x + 2 \ L \right) \right. 
\\
& + & \left. 
\left( { { \alpha_s } \over \pi } \right)^2  C_A^2 
 \left( - {2 \over 3} \ln^3 x + 2 \ln^2 x \ L - 2 \ln x \ L^2 
\right) + \ \cdots \right]   
\nonumber\\
\nonumber
\end{eqnarray}
The first two terms in the square bracket reproduce  respectively    
the leading-order result~\cite{loterm} and the 
$x \ll 1$ piece of the next-to-leading-order result~\cite{nloterm}; 
the third term determines the $x \ll 1$  pieces of the coefficients 
$c^{(2)}_{g g}$,  
${\overline c}^{(2)}_{g g}$,  
$ {\overline {\overline c}}^{(2)}_{g g}$ in  
Eq.~(\ref{coeffexp}). 

 The quark components of the coefficient 
 also contain 
 double-logarithmic corrections. These can be determined in an analogous 
 manner.  We find 
\begin{eqnarray} 
\label{cqg}
 C_{q g}  ( \alpha_s, x, {m_H^2 /  \mu^2} ) 
&=& 
{ { \alpha_s^2 m_H^2 G_F} \over { 288 \pi \sqrt{2} }}  
\left[  
{ {\alpha_s }\over \pi } \ 
C_F \left( -  \ln x +  L \right) 
+ \left( { { \alpha_s } \over \pi } \right)^2  C_F  C_A  
\hspace*{0.4 cm} \right. 
\nonumber\\
& \times & \left. 
 \left( - {1 \over 2} \ln^3 x + {3 \over 2} \ln^2 x \ L - 
{3 \over 2} \ln x \ L^2 
\right) + \ \cdots \right]    \hspace*{0.4 cm} , 
\\
\nonumber
\end{eqnarray} 
\begin{eqnarray} 
\label{cqq}
 C_{q {\bar q}} ( \alpha_s, x, {m_H^2 /  \mu^2} ) 
&=& 
{ { \alpha_s^2 m_H^2 G_F} \over { 288 \pi \sqrt{2} }}  
\\
& \times &
\left( { { \alpha_s } \over \pi } \right)^2 \ C_F^2 
\ \left( - {1 \over 3} \ln^3 x +  \ln^2 x \ L - 
 \ln x \ L^2 
\right) + \ \cdots 
\nonumber\\
\nonumber
\end{eqnarray} 
 
In summary, perturbative corrections to Higgs  production 
are double-logarithmic 
for $x \ll 1$ in the $m_t \to \infty$ effective theory, 
while single-logarithmic in the full theory. We have 
given  simple formulas 
that incorporate all the double logarithms 
to any order in $\alpha_s$ in the $m_t \to \infty$ case. 
The 
 formulas also contain 
subleading $N \to 0$ terms, mostly of kinematical origin, 
which  may be relevant 
for matching the $x \ll 1$ contributions 
with expansions around the soft 
limit $x \to 1$~\cite{hk02,cat02,soft01}. 
For observables sensitive to the $x \ll 1$ region, it will 
 be relevant to include finite top-mass  corrections.

\end{document}